\documentclass[final,5p,times,twocolumn]{elsarticle}

\usepackage{bm}
\usepackage{graphicx}
\usepackage{amssymb}
\usepackage{amsmath}
\usepackage{color}
\usepackage[utf8]{inputenc}
\usepackage[unicode=true,colorlinks=true,urlcolor=blue,citecolor=blue]{hyperref}
\usepackage{subfigure}

\usepackage{pifont}
\usepackage{ulem}
\usepackage[dvipsnames]{xcolor}

\usepackage{dcolumn}
\usepackage{bm}
\usepackage{hyperref}

\usepackage{orcidlink}
\usepackage{amsthm}
\usepackage{lineno}

\journal{Physica B: Condensed Matter }

\begin{document}
\let\WriteBookmarks\relax
\def\floatpagepagefraction{1}
\def\textpagefraction{.001}
\begin{frontmatter}



\title{Effect of spin-dependent tunneling and intervalley scattering in magnetic-semiconductor van der Waals heterostructures on exciton and trion polarization}

\author[inst1]{V.~N.~Mantsevich~\orcidlink{0000-0003-2879-153X}}
\fnmark[1]
\ead{vmantsev@gmail.com}

\affiliation[inst1]{organization={Lomonosov Moscow State University},
                city={Moscow},
                postcode={119991},
                country={Russia}}

\begin{abstract}
We present a theoretical analysis of valley pseudospin control in the transition metal dichalcogenide (TMD) monolayer by utilizing the
magnetic proximity effect of 2D magnetic layer and, propose self-consistent analysis of photoluminescence (PL) polarization peculiarities in TMD/magnetic material van der Waals heterostructures. We attribute observed peculiarities to the interplay between spin-dependent interlayer charge transfer and intervalley scattering of excitons and trions. The ratio between the electron tunneling timescale and the exciton and trion intervalley scattering lifetimes and radiative lifetimes determine the PL dynamics. A possibility to switch PL polarization sign due to the quasi-particles dynamics  under circularly polarized laser excitations is revealed. We also discuss generalization of the proposed model due to the careful analysis of both intervalley and intravalley scattering processes between bright and dark excitons. Obtained results allow a long-distance manipulation of exciton and trion behaviors and open the possibilities for the effective control under spin and valley pseudospin in multilayer magnetic-semiconductor van der Waals heterostructures.
\end{abstract}



\end{frontmatter}

\section{Introduction}
In recent years two-dimensional (2D) materials and van der Waals heterostructures \cite{Geim2013,Sierra2021} consisting of different 2D crystals have been among the most active areas of both theoretical and experimental research in solid-state physics due to their unique physical properties and promissing applications in ultrafast and low-power nanoelectronic and spintronic devices. 
Transition metal dichalcogenides (TMDs) have recently received special attention among 2D semiconductor materials due to their outstanding optical and electronic properties \cite{Manzeli2017}. Strong spin-orbit interaction in addition to direct band gap at the $K$-point of the Brillouin zone in the monolayer limit leads to effective spin-valley coupling. Thus, optical transitions in a specific valley can be selectively excited by circularly polarized light \cite{Mak2012,Ye2016}. This allows control of valley degree of freedom by optical means (helicity resolved photoluminescence measurements \cite{Zeng2012}), providing an important potential for processing information stored in spin/valley degree of freedom. Depending on the specific TMD material the mechanism for generating a valley polarization by means of circularly polarized excitation can be very effective and robust even in the case of non-resonant excitation.

However, this potential of TMDs van der Waals heterostructures has encountered practical challenges due to the dominant role of excitons in optical transitions \cite{Wang2018}. Unfortunately, the valley polarization of excitons has a very short lifetime on the orders of a few picoseconds \cite{Ye2016} making exciton-based valleytronics impractical. Excitonic valley polarization can be achieved by lifting the energy degeneracy of opposite valleys, which is usually done by means of external magnetic fields applied perpendicularly to the plane of TMD \cite{MacNeill2015}, but external magnetic fields should be rather strong, that does not fit for device application. An alternative strategy previously adopted for traditional semiconductors deals with the use of proximity effect \cite{Maslova2018,Mantsevich2019}. In this case a magnetic material stacked with the semiconductor induces magnetization due to the exchange interaction \cite{Scharf2017,Zhong2017}. In this context, the discovery of 2D magnets realizes a huge leap for the appearance of wholly van der Waals based spin(valley)tronics \cite{Huang2018,Gong2019,Burch2018}. The emergence of ferromagnetism in 2D van der Waals crystals in addition to their unique electronic and optical properties could open wide possibilities for 2D magnetic, magneto-electric and magneto-optic applications \cite{Burch2018,Gibertini2019}. Among the first discovered 2D magnetic structures are widely studied  $Cr_{2}Ge_{2}Te_{6}$ (metal) \cite{Gong2017} and $CrI_{3}$ (semiconductor) \cite{Sivadas2015,Zhang2015} crystals. Many other notable examples of magnetic order in 2D are obtained by molecular beam epitaxial growth of $FePS_{3}$ \cite{Wang2016}, $VSe_{2}$ \cite{Bonilla2018}, $MnSe_{2}$ \cite{OHara2018}, $Fe_{3}GeTe_{2}$ \cite{Fei2018}, and $CrX_{3}Te$ ($X=Si,Ge$) \cite{Li2014}. Later a control of magnetic order in $CrI_{3}$ bi-layers by means of electric fields was performed experimentally \cite{Huang2018}. At a fixed magnetic field near the metamagnetic transition a voltage-controlled switching between antiferromagnetic and ferromagnetic states was realized. Recently, the novel van der Waals antiferromagnet with ferromagnetic intralayer spin allignment $CrSBr$, has revealed stricking features of quasi-particle interactions \cite{Markina2026,Gser1990}. 

Van der Waals heterostructures based on 2D magnetic materials and TMDs monolayers are expected to combine the advantageous chiral optical selection rules and spin-valley locking of TMDs with the highly correlated and field-responsive long-range ordering inherent to magnets \cite{Xu2014,Zhong2017,Tong2019}.  In particular, the break of time-reversal symmetry resulting from the interaction between TMD and adjacent magnetic layers was demonstrated to induce spontaneous valley splitting and increased valley polarization, which manifest in modification of photoluminescence (PL) spectra \cite{Zhong2017,Choi2022}. These behaviors can also be achieved only due to the spontaneous magnetization in 2D magnetic materials \cite{Ciorciaro2020,Seyler2018}. While there have been a number of experimental studies among the plethora of available TMDs and 2D magnetics, all these structures reveal a common feature: an important role of spin-selective charge transfer between TMD monolayer and 2D magnetic material, which leads to a change in polarized PL intensity, such as in $WSe_2/CrI_3$, $MoSe_2/CrBr_3$, $MoSe_2/Cr_2Ge_2Te_6$, $MoSe_2/CrI_3$, and $MoSe_2/CrSBr$ 
heterostructures \cite{Zhong2020,Lyons2020,Zhang2022,Hong2025,Huang2024}. Similar effects were observed in more complicated systems consisting of TMD bilayer and 2D magnetic layer \cite{Wu2024} or structures with the intermediate layer of hexagonal boron nitride \cite{Dang2024}. In all these heterostructures strong spin polarization of 2D magnetic layer conduction bands gives rise to spin-dependent interlayer charge transfer for electrons, which results in strongly enhanced degree of PL circular polarization associated with excitons and trions recombination. 

The strong magnetic proximity effect at the TMD/magnetic material van der Waals heterointerfaces allows to analyze rich exciton and trion physics which is still not fully explored. One of the most important effects which directly influences the PL properties of van der Waals heterostructures in addition to spin-dependent charge transfer between the layers is the intervalley scattering of excitons and trions \cite{Zhang2022,Kioseoglou2012,Surrente2018} which strongly influence quasi-particles lifetime.  Several studies proposed that the electron-hole exchange interaction activates the intervalley scattering \cite{Mai2013,Yu2014}. However, the others debate that the electron-phonon scattering is the dominant mechanism \cite{MolinaSnchez2017,Carvalho2017,Miller2019}. In general, spin-valley exciton dynamics is collectively governed by electron-phonon scattering, spin-orbit interaction, and electron-hole interactions \cite{Jiang2021}. So, to correctly describe PL dynamics of excitons and trions in TMD/magnetic material van der Waals heterointerfaces one should carefully consider the interplay between spin-dependent charge transfer between the semiconductor and magnetic layers and the intervalley scattering of excitons and trions. Despite a significant amount of experimental data, no consistent theoretical model describing the role of both the interlayer spin dependent charge transfer and intervalley scattering of excitons and trions has been proposed to date. 

In this paper we propose a general theory which considers both the interlayer spin dependent charge transfer and intervalley scattering of excitons and trions and perfectly describes the peculiarities of PL polarization in TMD/magnetic material van der Waals heterostructures. The proposed model allows to analyze microscopically the problem of long-distance manipulation of bright exciton and trion behaviours in multilayer TMD/magnetic heterostructures and includes all important features of quasi-particles dynamics. The model operates with parameters directly available from experimental measurements and suits for a wide class of heterostructures with a similar design.

\section{Theoretical model}

\begin{figure}[h]
  \centering
  \includegraphics[width=0.95\linewidth]{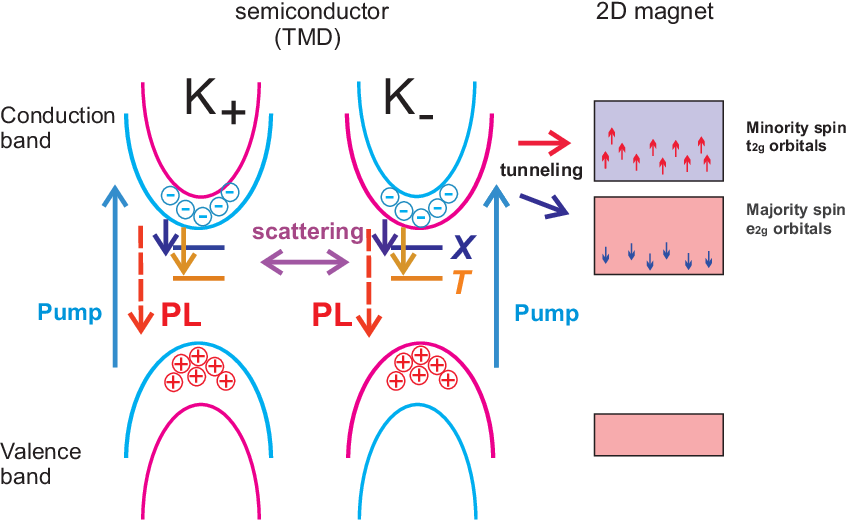}
  \caption{Scheme of the processes causing the population dynamics under laser excitation (absorption, formation of excitons and trions, emission, spin-dependent interlayer charge transfer over the 2D magnet-semiconductor interface and intervalley scattering). Semiconductor part demonstrates the sketch of TMD conduction and valence bands at $K_{\pm}$ points of the Brillouin zone, 2D magnet part depicts band structure of the magnetic layer (rectangles correspond to the 2D magnet band structure and show two spin-split electron sub-bands). Excitation occurs by means of laser pumping, photoluminescence signal arises due to the recombination in the semiconductor. Non-radiative channel corresponds to the tunneling channels between semiconductor and 2D magnet conduction bands with different tunneling rates shown by red and black curves (resonant and non-resonant tunneling). The model considers processes of exciton and trion intervalley scattering (magenta horizontal arrow). }
  \label{fig:scheme}
\end{figure}

Being motivated by the experiments~\cite{Lyons2020,Hong2025,Wu2024}, we consider a general theoretical approach and do not restrict ourselves by a particular design of TMD/magnetic material van der Waals heterostructure. The model system under investigation is shown in Fig.~\ref{fig:scheme}. The main feature of the considered model is a simultaneous consideration of two effects affecting the dynamics of PL: interlayer spin dependent charge transfer through the TMD/magnetic material van der Waals heterointerface \cite{Lyons2020,Hong2025,Wu2024,Zhang2022} and intervalley scattering of bright excitons and trions \cite{Zhang2022,Kioseoglou2012,Surrente2018,BergmannIwe2025}. We would like to note that parameters of spin-dependent transfer strongly depend on the magnetic state of the 2D magnet \cite{Wilson2021,SeratideBrito2023} and intervalley scattering significantly influence the lifetime of quasi-particles. So, the interplay between the typical time scales present in the system determines the dynamics of measured PL and will be carefully studied in the present paper. One more important effect, which should be taken into account deals with the presence of localized excitons in addition to neutral bright excitons and trions. Localized excitons are confined by defect potentials originating from the magnetic layer \cite{Huang2024}. Such excitons were detected in vertically stacked $MoSe_2/MnPSe_3$, $MoSe_2/FePSe_3$ and $MoSe_2/Cr_2Ge_2Te_6$ due to the appearance of new peaks between the designated axciton and trion peaks \cite{Onga2020}.

At the initial time moment $t=0$ 2D electrons in the TMD are excited by means of linearly or circularly polarized non-resonant optical pumping with the photon energy larger than the band gap of 2D semiconductor. An important feature of both experimental measurements and the proposed theory is that photoexcitation occurs only in TMD and no electron-hole pairs are created in the 2D magnetic layer. Experimentally, this condition can be fulfilled by performing excitation with photon energy which corresponds to the absorption minimum of the 2D magnetic material \cite{Lyons2020,Bermudez1979,Dillon1966}. Shortly after excitation a part of electrons and holes could become bound into pair states due to the presence of Coulomb interaction (bright exciton) with the bound rate $\gamma_{X}$. Further time evolution of excited quasi-particles could lead to the formation of {positive or negative trions} with the bound rate $\gamma_{T}$ and localized excitons with the bound rate $\gamma_{loc}$. All these bound rates can be determined directly from the experimental measurements. The formation of bright excitons $X$ and trions $T$ is shown in Fig.\ref{fig:scheme} by vertical black arrows. Localized excitons appear due to the capture of excitons by deep traps \cite{Terres2025,Rabouw2015}. We do not consider detrapping processes, so localized excitons can only recombine with typical lifetime being larger that the lifetime of free excitons and trions \cite{Huang2024,Terres2025,Kurilovich2020,Kurilovich2022}. An important experimental fact is that typically due to the heterostructure band alignment the observed trion luminescence is attributed to {positive trions} \cite{Lyons2020,SeratideBrito2023}. We will consider in our model only this type of trions as negative trions do not contribute to the circular polarization of PL, as they are formed by a single hole and two electrons with opposite spins.

A spin-split sub-bands of the 2D magnetic material are separated from the TMD layer by a tunnel barrier, which is different for spin-up and spin-down electrons tunneling from the TMD layer to the 2D magnetic layer. Due to the type-II interlayer band alignment in multilayer TMD/magnetic heterostructures two channels for electron with opposite spins tunneling are present (see Fig.\ref{fig:scheme} for details). Both channels are generally  momentum-nonconserving due to momentum mistmatch between the corresponding minima of the conduction bands in the TMD and magnetic material, but an important fact is that tunneling through one of the channels will be more efficient than through the another. This difference is the key point of the proposed model and for simplicity, further one of the channels will be called "resonant" and another one - "non-resonant" following the results obtained in \cite{Lyons2020}. 

To model the dynamics of electrons, excitons (free and localized) and trions and to carefully explain experimental results on the observation of circular polarization of exciton and trion PL, we propose a theoretical model which describes the time evolution of the occupancies of spin-up and spin-down electrons $n_{k\uparrow(\downarrow)}$, free $N_{k\uparrow(\downarrow)}^{X}$ and localized $N_{k\uparrow(\downarrow)}^{Xloc}$ excitons and positive trions $N_{k\uparrow(\downarrow)}^{T}$. We neglect the tunneling of holes, as typically photogenerated holes are prohibited to tunnel between the layers due to the rather large effective mass and large valence band offset \cite{Zhang2022}. Generation of spin-up and spin-down electrons is the initial stage of further complex dynamics of the system. It is described by the term $g_{\uparrow(\downarrow)}$. Spin dependent tunneling of electrons between TMD and 2D magnetic layer interface is described by the tunneling rate $\Gamma_{\sigma}(k)=\pi\nu_{0}(\varepsilon_{k})t_{k}^{2}$ with index $\sigma=\uparrow,\downarrow$. $t_{k}$ is the tunneling amplitude determined by the overlapping of electron wave functions in the 2D magnetic layer and in the TMD; $\nu_{0}(\varepsilon_{k})$ is the unperturbed density of states in the TMD. Furthermore, we assume that the tunneling parameter $t_{k}$ has a negligibly weak dependence on $k$, so for 2D density of states in the TMD the tunneling rate is a constant $\Gamma_{\sigma}(k)=\Gamma_{\sigma}$, which we take as a parameter determined directly from experimental measurements. Exciton and trion intervalley scattering processes are described by the scattering times $T_{sc}^{X}$ and $T_{sc}^{T}$, which are also known from the experimental measurements \cite{Zhang2022,Kioseoglou2012,Surrente2018,BergmannIwe2025}.  We will neglect nonlinear processes, such as exciton-exciton interaction as it becomes important in the experiments performed with high exciton density \cite{Steinhoff2024,Kumar2014,Zhao2003,Adejumobi2024,Maslova2025}. This processes does not strongly influence the dynamics of quasi-particles and can be easily considered by adding a non-liner term in equation describing exciton dynamics \cite{Kurilovich2024}. Finally, we will consider that excitons (free and localized) and trions have finite lifetimes  $\tau_{X}$, $\tau_{loc}$ and $\tau_{T}$, respectively. We will analyze the situation, when all the time scales are strongly shorter than the electron spin relaxation time $\tau_{0}$ that is about 10 ns in 2D structures \cite{Korenev2012}. Semiconductor material could be doped, so charge carriers could be present even in the absence of laser excitation. Let us consider a situation of a p-doped TMD monolayer, when resident holes are present \cite{Lyons2020,Huang2024,Hong2025}. So, occupancies of electrons, free and localized excitons and trions satisfy the set of rate equations
\begin{eqnarray}
&&\frac{dn_{k\sigma}}{dt}=g_{\sigma}(1-n_{k\sigma})\Theta(t)-\Gamma_{\sigma}n_{k\sigma}\nonumber\\&&-\gamma_{X}n_{k\sigma}(1+N^{X}_{k\sigma})-\gamma_{T}n_{k\sigma}(1-N^{T}_{k\sigma})-\frac{n_{k\sigma}}{\tau_{0}},\nonumber\\
&&\frac{dN^{T}_{k\sigma}}{dt}=\gamma_{T}n_{k\sigma}(1-N^{T}_{k\sigma})+\gamma_{T}N^{X}_{k\sigma}(1-N^{T}_{k\sigma})\nonumber\\&&\mp\frac{N^{T}_{k\sigma}-N^{T}_{k-\sigma}}{T_{sc}^{T}}-\frac{N^{T}_{k\sigma}}{\tau_{T}},\nonumber\\
&&\frac{dN^{X}_{k\sigma}}{dt}=\gamma_{X}n_{k\sigma}(1+N^{X}_{k\sigma})-\gamma_{T}N^{X}_{k\sigma}(1-N^{T}_{k\sigma})\nonumber\\&&\mp\frac{N^{X}_{k\sigma}-N^{X}_{k-\sigma}}{T_{sc}^{X}}-\gamma_{loc}N^{X}_{k\sigma}-\frac{N^{X}_{k\sigma}}{\tau_{X}},\nonumber\\
&&\frac{dN^{Xloc}_{k\sigma}}{dt}=\gamma_{loc}N^{X}_{k\sigma}-\frac{N^{Xloc}_{k\sigma}}{\tau_{loc}},
\label{eqn:system}
\end{eqnarray}
An important advantage of the proposed model is the absence of fitting parameters, as all the quantities, which determine the dynamics of quasi-particles and, consequently, the dynamics of PL can be directly determined from the experimental measurements performed under particular materials and the geometry of heterostructure. 

The spin polarization of excitons and trions remaining in the TMD layer is given by $N_{\uparrow}^{X(T)}-N_{\downarrow}^{X(T)}$, where $N_{\sigma}^{X(T)}=\int N_{k\sigma}^{X(T)}(\varepsilon_{k})d\varepsilon_{k}$ can be analyzed experimentally by means of measuring the degree of circular polarization of the exciton and trion PL as
\begin{eqnarray}
P_{X(T)}=\frac{N_{\uparrow}^{X(T)}-N_{\downarrow}^{X(T)}}{N_{\uparrow}^{X(T)}+N_{\downarrow}^{X(T)}}.
\end{eqnarray}
Further we will analyze the exciton and trion PL circular polarization time evolution due to the interplay between interlayer spin dependent charge transfer and intervalley scattering of free excitons and trions. 
 
\section{Results and discussion}

\subsection{Lineraly polarized light excitation and fixed exciton and trion intervalley scattering rates}

\begin{figure} [h]
  \centering
  \includegraphics[width=\linewidth]{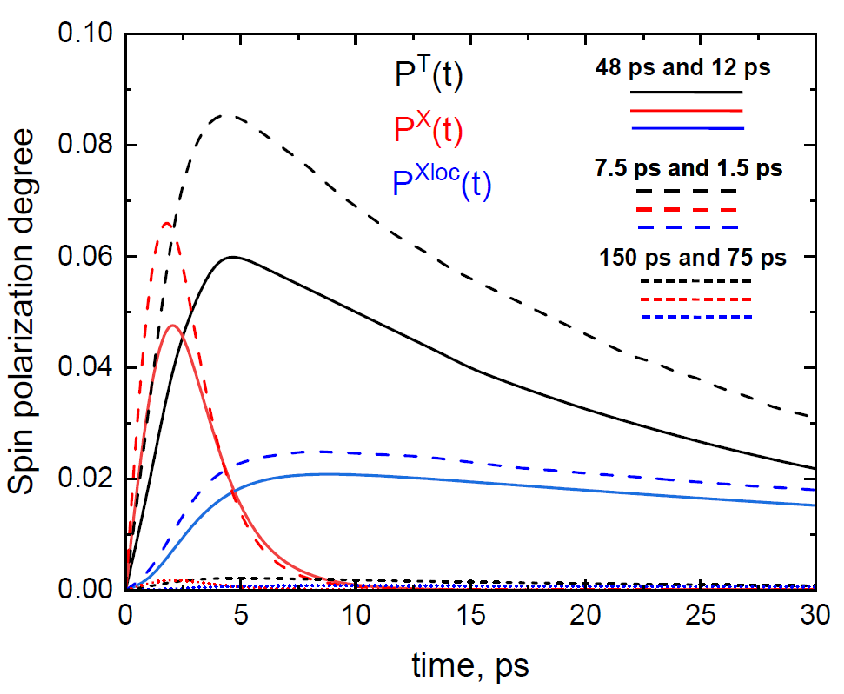}
  \caption{Spin polarization degree time evolution for excitons (red curves), trions (black curves) and localized excitons (blue curves) under the linearly polarized light excitation. Different electron interlayer tunneling rates are considered and exciton and trion intervalley scattering rates are fixed. Solid curves correspond to the situation, when both tunneling rates of electrons are faster than the free exciton lifetime but slower than the trion and localized exciton lifetimes. Dashed curves demonstrate results in the case when both tunneling rates of electrons are faster than free and localized exciton and trion lifetimes. Short dashed curves reveal results in the case when both electron tunneling rates are slower than free and localized exciton and trion lifetimes. The following system parameters were used: $\tau_{X}=10$ ps, $\tau_{T}=50$ ps and $\tau_{loc}=60$ ps (see \cite{Zipfel2022,Venanzi2024,Rabouw2015,Terres2025}); $\gamma_{X}^{-1}=2$ ps, $\gamma_{T}^{-1}=5$ ps and  $\gamma_{loc}^{-1}=10$ ps (see \cite{Venanzi2024,Mourzidis2025,Ceballos2016,Terres2025}); $T_{sc}^{X}=1.5$ ps and $T_{sc}^{T}=100$ ps (see \cite{Zhang2022}). Electron tunneling times $\Gamma_{\uparrow(\downarrow)}^{-1}$ are shown in the panel (see \cite{Lyons2020}).} 
  \label{fig:modelling}
\end{figure}

Let us first discuss the situation when excitation occurs by means of linearly polarized light. In this case unpolarized nonequilibrium 2D electrons with the energies
$\varepsilon_{k}$, where $k$ is the in-plane wave vector are excited in both valleys of TMD monolayer, as the concentration of spin-up and spin-down electrons in this case is nearly the same. The peculiarities of the exciton and trion PL circular polarization in this case are governed by the ratio between timescale of electron spin-dependent tunneling transfer, the exciton and trion intervalley scattering times and their radiative lifetimes. To perform careful modelling of quasi-particles dynamics and time evolution of the PL polarization we used the system parameters determined directly from the experimental measurements performed under the TMD/magnetic heterostructures. Similar behavior of PL circular polarization was experimentally observed for different TMDs and 2D magnetic materials. So, our model allows to describe a large variety of TMD/magnetic heterostructures, one should only use the parameters corresponding to the system under study. To be concrete we will focus on the $MoSe_2$ and $MoS_{2}$ based heterostructures. In our modelling we used the following system parameters: exciton and trion lifetimes were considered to be about $\tau_{X}=10$ ps and $\tau_{T}=50$ ps, which are the averaged values observed in $MoSe_2$ \cite{Zipfel2022,Venanzi2024} and $MoS_2$ \cite{Wang2016}, typical times of exciton and trion formation are about $\gamma_{X}^{-1}=2$ ps and $\gamma_{T}^{-1}=5$ ps \cite{Mak2012,Zhang2022,Venanzi2024,Mourzidis2025,Ceballos2016}. Localized excitons are formed longer than trions as they should be captured by deep traps. Trapping times can be estimated to be about $\gamma_{loc}^{-1}=10-100$ ps \cite{Terres2025}. Lifetime of localized excitons also exceeds lifetime of free excitons and is about the lifetime of trions or even longer, so it is about $\tau_{loc}=50-500$ ps \cite{Rabouw2015,Terres2025}. Finally, free exciton and trion intervalley scattering times are in the range $T_{sc}^{X}=1-4$ ps and $T_{sc}^{T}=50-150$ ps, correspondingly \cite{Zhang2022,Mai2013,Cao2012,Lagarde2014}. Localized excitons do not experience intervalley scattering. We also considered the electron tunneling rates to change in the following range $\Gamma_{\uparrow(\downarrow)}^{-1}=2-120$ ps. We note that the tunneling rates may vary over a large range, as was revealed in recent pump-probe experiments on TMD based heterostructures \cite{Merkl2019}. Electron tunneling rates in TMD heterostructures vary greatly, driven by band alignment (Type-I vs. Type-II), layer thickness, carrier density, and interface quality \cite{Das2014}. Moreover, one can tune the tunneling rates by means of the external gate voltage changing, which is a typical way for tunneling characteristics control in heterostructures \cite{Sierra2021}. Calculation results for different ratios between tunneling timescale, intervalley scattering times and radiative lifetimes are shown in Fig.\ref{fig:modelling}.

We will first study the situation, when free exciton and trion intervalley scattering rates continue being constant and focus on the quasi-particles dynamics due to the different ratio between electron interlayer tunneling rates and quasi-particles lifetimes. Obtained results demonstrate that polarization of free and localized excitons and trions depend on the ratio between the timescales present in the system. As formation of quasi-particles occurs sequentially, the PL polarization corresponding to different types of quasi-particles also appear at a particular time moment: free exciton polarization appears faster than the trion polarization and localized exciton polarization because free exciton formation times are shorter than the trion and localized excitons formation times. The slowest process is the formation of localized excitons polarization. The polarization of both free and localized excitons and trions is nearly absent in the case when tunneling rates of spin-up and spin-down electrons are slower than the exciton and trion lifetimes (see short dashed lines in Fig.\ref{fig:modelling}). In this case electrons with opposite spins do not have enough time to tunnel to the 2D magnetic layer and the imbalance of carriers with opposite spins does not have enough time to manifest itself in the PL polarization. So, the polarization of carriers themself and, consequently, circular polarization of PL are nearly absent. The opposite situation, when both spin-up and spin-down electron tunneling rates are faster than exciton and trion lifetimes is show by the dashed curves. In this case polarization of excitons and trions is well pronounced but it corresponds to the situation, when the concentration of electrons which could tunnel resonantly (faster) is nearly absent and a very small amount of electrons with opposite spin, which tunnel non-resonantly (slower) are present in the system. Spin polarization of free excitons decays rather fast and a well pronounced decay of trion polarization and localized excitons polarization can be seen. Unfortunately, it is rather difficult to detect PL signal in this regime due to its weak amplitude caused by the very low concentration of quasi-particles, which could effectively recombine. The most perspective situation from experimental point of view is shown by the solid curves. In this case tunneling of both electrons with opposite spins is faster than the trion and localized excitons lifetime but slower than the free exciton lifetime. In this regime free exciton polarization dominates at short times and localized exciton and trion polarization become dominant at longer times and continue being well resolved for a long time. Trion PL continue being larger than the localized exciton polarization during the whole period of PL time evolution. Both sub-populations of spin-up and spin-down electrons have enough concentration to detect circular polarization of PL experimentally \cite{Lyons2020}. So, one can reveal PL circular polarization associated mostly with free excitons, trions or localized excitons, or even all of the quasi-particles together by changing the time moment for PL measurements. 

\subsection{Lineraly polarized light excitation and fixed electron interlayer tunneling rates}

\begin{figure} [h]
  \centering
  \includegraphics[width=\linewidth]{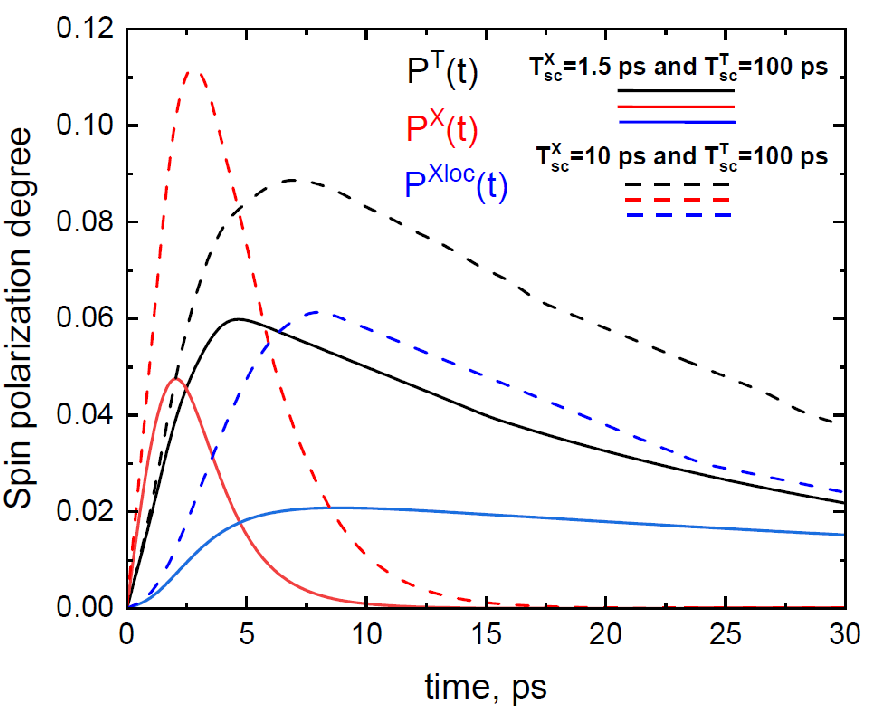}
  \caption{Spin polarization degree time evolution for excitons (red curves), trions (black curves) and localized excitons (blue curves) under the linearly polarized light excitation. Different exciton and trion intervalley scattering rates are considered and electron interlayer tunneling rates are fixed. Solid curves correspond to the situation, when fast intervalley scattering takes place. Dashed curves demonstrate results in the case when slow intervalley scattering is present. The following system parameters were used: $\tau_{X}=10$ ps, $\tau_{T}=50$ ps and $\tau_{loc}=60$ ps (see \cite{Zipfel2022,Venanzi2024,Rabouw2015,Terres2025}); $\gamma_{X}^{-1}=2$ ps, $\gamma_{T}^{-1}=5$ ps and  $\gamma_{loc}^{-1}=10$ ps (see \cite{Venanzi2024,Mourzidis2025,Ceballos2016,Terres2025}); $\Gamma_{\uparrow}^{-1}=48$ ps and $\Gamma_{\downarrow}^{-1}=12$ (see \cite{Lyons2020}). Exciton and trion intervalley scattering times are shown in the panel (see \cite{Zhang2022}).} 
  \label{fig:modelling_2}
\end{figure}

Let us now reveal the role of intervalley scattering processes. We will fix the electron interlayer tunneling rates considering the situation, when tunneling of both electrons with opposite spins is faster than the trion and localized exciton lifetimes but slower than the free exciton lifetime, and study the PL behavior for different exciton and trion intervalley scattering rates. Typical inteervalley scattering times for trions are rather large \cite{Zhang2022} in comparison with all other system time scales, so they do not strongly influence the dynamics of the PL and we will fix this parameter. Meanwhile, exciton intervalley scattering times are rather fast \cite{Zhang2022} and changing this timescale could lead to a visible modification of PL polarization. Let us compare two opposite situations, when fast and slow intervalley exciton scattering occurs. Obtained results are shown in Fig.\ref{fig:modelling_2}. Solid curves correspond to the situation, when fast intervalley scattering takes place. Dashed curves demonstrate results in the case when slow intervalley scattering is present. It is clearly evident, that fast intervalley scattering causes the decreasing of PL circular polarization degree for all the quasiparticles, as it 
suppresses valley polarization of excitons. The damping of free excitons and, consequently, localized excitons PL polarization with the increasing of exciton intervalley scattering (decreasing of intervalley scattering timed) is much larger than for trions, because the valley-polarization lifetime of trions has been proven to be much longer than of excitons. To conclude, the presence of intervalley scattering opposes the growth of PL circular polarization formed due to the electron intervalley tunneling. It always suppresses the degree of polarization and as it is rather fast in the considered materials, its influence is well pronounced, contrary to the interlayer tunneling, which could lead to a significant growth of PL circular polarization.

\subsection{Circularly polarized light excitation}

\begin{figure} [h]
  \centering
  \includegraphics[width=\linewidth]{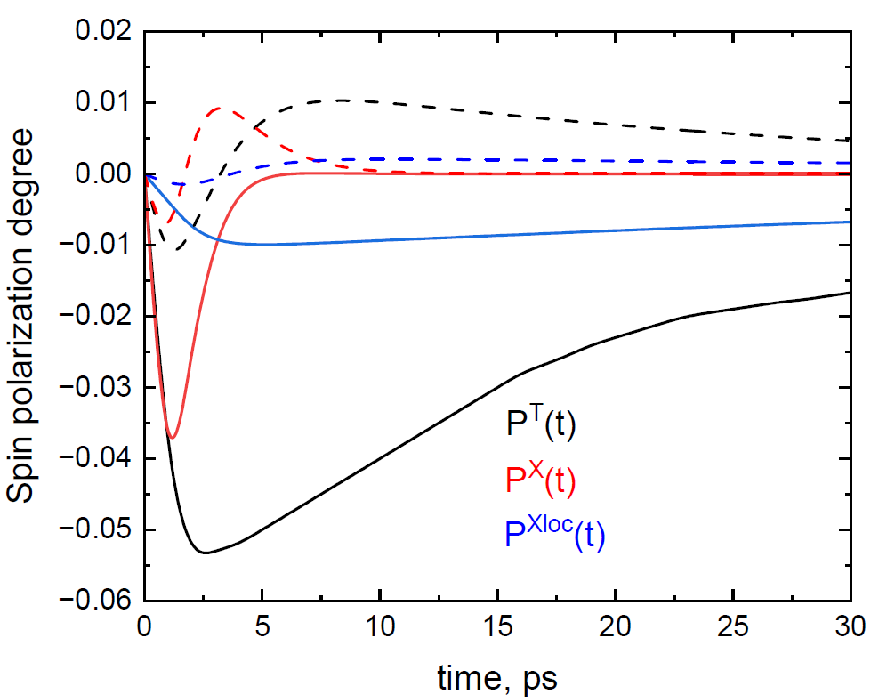}
  \caption{Spin polarization degree sign changing (solid curves) and sign switching (dashed curves) for excitons (red curves), trions (black curves) and localized excitons (blue curves). The system parameters are the same as for solid curves in Fig.\ref{fig:modelling}. The initial polarization of electrons due to the excitation by means of circularly polarized light is ten percent for solid curves and five percent for dashed curves.  }
  \label{fig:sign_switching}
\end{figure}

Let us now analyze the dynamics of PL circular polarization under the excitation by means of circularly polarized light. One could expect two effects: the PL circular polarization sign changing and switching of PL circular polarization sign. The implemented mode depends on the interplay between spin dependent charge transfer rate, intervalley exciton and trion scattering rate, quasi-particles lifetime and the degree of initial valley polarization. Let us analyze calculation results for the system parameters used for the solid curves in Fig.\ref{fig:modelling} and two different initial polarizations of electrons - ten percent polarization and five percent polarization. Obtained results are shown in Fig.\ref{fig:sign_switching}. Let us first discuss the situation, when switching of PL circular polarization sign occurs (see dashed curves in Fig.\ref{fig:sign_switching}) and initial circular polarization being five percent.

This is a net effect of the two: excitation of charge carriers in TMD monolayer by means of circular polarized light and the presence of fast and slow interlayer tunneling channels through the TMD/magnetic heterointerface. Excitation by means of circularly polarized light results in the imbalance of spin-up and spin-down electrons in the TMD monolayer due to the different concentration of excited carriers in two valleys. Switching of PL circular polarization sign appears in the situation, when the initial concentration of electrons which could tunnel fast through the heterointerface exceeds the concentration of electrons with the opposite spin projection which could tunnel slower (see dashed curves Fig.\ref{fig:sign_switching}). At the initial stage of the system time evolution polarization of PL is determined by the initial imbalance of spin-up and spin-down electrons in different valleys. During the system time evolution one can imagine the situation, when electrons with the opposite spin start to dominate due to the difference between tunneling rates. The majority of electrons tunnel faster than the minority, so the majority becomes a minority and vice versa. This directly means that PL sign starts to be determined by the electrons, whose concentration was smaller at the initial time moment, but at a certain time moment it dominates. So, circular polarization sign switching occurs both for free and localized excitons and trions associated PL (see Fig.\ref{fig:sign_switching}). It should be noted that for localized excitons this effect is negligibly small in comparison with free excitons and trions because electrons bound in localized excitons nearly could not effectively tunnel through the heterointerface. For the larger value of initial polarization (10 percent, see solid curves Fig.\ref{fig:sign_switching}) this effect is not so well pronounced and PL circular polarization sign switching does not occur due to the system timescales. For larger difference between spin-up and spin-down electrons concentration spin-dependent tunneling through the heterointerface does not have enough time to change the concentration ratio. It only changes the sign of PL circular polarization to the opposite one in comparison with the results shown in Fig.\ref{fig:modelling}.

\subsection{Generalization of the proposed model. Exciton dynamics channels and the role of correlated mechanisms.}

In the previous section we have performed modeling for a specific system and a specific type of materials, but it should be noted, that our model is rather general and can be applied to a wide class of van der Waals heterostructures and consider a number of more subtle effects. These additional patterns will not undergo qualitative changes, as physical mechanism is nearly the same, but only quantitative ones. First of all in the case of another types of materials and samples structure (an amount of different materials layers \cite{Zhao2025}, layers order  \cite{Wu2024,Dang2024}, direction of carriers tunneling \cite{Zhang2022,Dang2024}) one should use different system parameters (tunneling rates, lifetimes, times of quasi-particles formation) \cite{Zhu2014}. The proposed model could also implement the role of external magnetic field, which could lead to the controllable switching of the 2D magnetic material state from anti-ferromagnetic to ferromagnetic \cite{Huang2018}, leading to the modification of tunneling channels characteristics between semiconductor and magnet layers. One more possible generalization of the proposed model deals with the regime of high intensity excitation of TMD/magnetic heterostructure. In this regime high exciton density leads to the need to consider exciton-exciton interaction \cite{Wietek2024,Kurilovich2024}. In this case one should add the terms $-R_{A}[(N^{X}_{k\sigma})^{2}+N^{X}_{k\sigma}N^{X}_{k-\sigma}]$ to the equations describing exciton dynamics in Eqn. (\ref{eqn:system}), where $R_{A}$ is the constant of exciton-exciton interaction. It can be determined directly from experimental measurements \cite{Wietek2024}. 

\begin{figure}[h]
  \centering
  \includegraphics[width=0.85\linewidth]{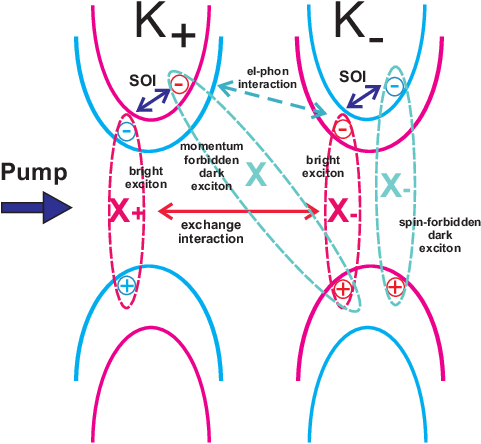}
  \caption{The general scheme of the intervalley and intravalley bright exciton scattering. The transitions occur due to the electro-hole exchange interaction, spin-orbital interaction and electron-phonon interaction.  }
  \label{fig:scattering}
\end{figure}

Finally, one could consider the appearance of different types of excitons. In addition to direct bright excitons intervalley scattering due to the electron-hole exchange interaction or electron-phonon scattering discussed above one could take into account intravelley scattering processes \cite{Jiang2021}. This processes will not qualitatively influence the dynamics of PL, but could lead to quantitative changes, which could be important for experimental measurements. A bright exciton $N^{Xbr}_{k\sigma}$ can transfer into optically inactive spin-forbidden dark exciton $N^{Xsf}_{k\sigma}$ through the spin flip of the electron or hole, which are realized through an external magnetic field or internal spin-orbit interaction. It may also transfer to optically inactive a momentum-forbidden dark exciton $N^{Xmf}_{k\sigma}$ with the electron and hole located in different valleys through the scattering with defects or phonons. The scheme of the pathways is shown in Fig.\ref{fig:scattering}. 

\begin{figure} [h]
  \centering
  \includegraphics[width=\linewidth]{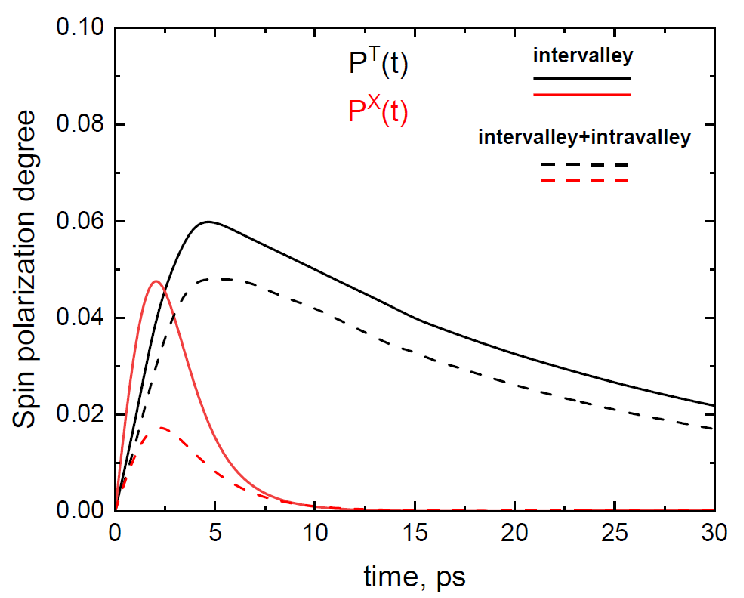}
  \caption{A comparison of spin polarization degree time evolution for bright excitons (red curves) and trions (black curves) under the linearly polarized light excitation. Solid curves demonstrate results when only intervalley scattering processes are considered. Dashed curves correspond to the situation when both intervalley and intravalley scattering between bright and dark excitons are taken into account. Calculation parameters are the same as were used in Fig.\ref{fig:modelling}. The following system parameters were used: $\tau_{X}=10$ ps, $\tau_{T}=50$ ps and $\tau_{loc}=60$ ps (see \cite{Zipfel2022,Venanzi2024,Rabouw2015,Terres2025}); $\gamma_{X}^{-1}=2$ ps, $\gamma_{T}^{-1}=5$ ps and  $\gamma_{loc}^{-1}=10$ ps (see \cite{Venanzi2024,Mourzidis2025,Ceballos2016,Terres2025}); $T_{sc}^{X}=1.5$ ps and $T_{sc}^{T}=100$ ps (see \cite{Zhang2022}); $\Gamma_{\uparrow}^{-1}=48$ ps and $\Gamma_{\downarrow}^{-1}=12$ (see \cite{Lyons2020}); $T_{SOI}^{X}=2$ ps and $T_{el-phon}^{X}=50$ fs (see \cite{Jiang2021,Bertoni2016}).} 
  \label{fig:comparison}
\end{figure}

Following the scheme, one can modify system of Eqs.(\ref{eqn:system}) and analyze the PL dynamics of bright excitons taking carefully into account all possible transitions between different types of excitons existing in the heterostructure. Trions and localized excitons are formed directly from the bright excitons as it was considered previously. The modified system of equation reads

\begin{eqnarray}
&&\frac{dn_{k\sigma}}{dt}=g_{\sigma}(1-n_{k\sigma})\Theta(t)-\Gamma_{\sigma}n_{k\sigma}\nonumber\\&&-\gamma_{X}n_{k\sigma}(1+N^{Xbr}_{k\sigma})-\gamma_{T}n_{k\sigma}(1-N^{T}_{k\sigma})-\frac{n_{k\sigma}}{\tau_{0}},\nonumber\\
&&\frac{dN^{T}_{k\sigma}}{dt}=\gamma_{T}n_{k\sigma}(1-N^{T}_{k\sigma})+\gamma_{T}N^{Xbr}_{k\sigma}(1-N^{T}_{k\sigma})\nonumber\\&&\mp\frac{N^{T}_{k\sigma}-N^{T}_{k-\sigma}}{T_{sc}^{T}}-\frac{N^{T}_{k\sigma}}{\tau_{T}},\nonumber\\
&&\frac{dN^{Xbr}_{k\sigma}}{dt}=\gamma_{X}n_{k\sigma}(1+N^{Xbr}_{k\sigma})-\gamma_{T}N^{Xbr}_{k\sigma}(1-N^{T}_{k\sigma})\nonumber\\&&\mp\frac{N^{Xbr}_{k\sigma}-N^{Xbr}_{k-\sigma}}{T_{sc}^{X}}-\frac{N^{Xbr}_{k\sigma}-\alpha N^{Xsf}_{k-\sigma}}{T_{SOI}^{X}}-\frac{N^{Xbr}_{k\sigma}-\beta N^{Xmf}_{k\sigma}}{T_{el-phon}^{X}}\nonumber\\&&
-\gamma_{loc}N^{Xbr}_{k\sigma}-\frac{N^{Xbr}_{k\sigma}}{\tau_{X}},\nonumber\\
&&\frac{dN^{Xsf}_{k\sigma}}{dt}=\frac{N^{Xbr}_{k\sigma}-\alpha N^{Xbr}_{k-\sigma}}{T_{SOI}^{X}}-\frac{N^{Xsf}_{k\sigma}}{\tau_{sf}},\nonumber\\
&&\frac{dN^{Xmf}_{k\sigma}}{dt}=\frac{N^{Xbr}_{k\sigma}-\beta N^{Xmf}_{k\sigma}}{T_{el-phon}^{X}}+\frac{N^{Xbr}_{k-\sigma}-\alpha N^{Xmf}_{k\sigma}}{T_{SOI}^{X}}-\frac{N^{Xmf}_{k\sigma}}{\tau_{mf}},\nonumber\\
&&\frac{dN^{Xloc}_{k\sigma}}{dt}=\gamma_{loc}N^{Xbr}_{k\sigma}-\frac{N^{Xloc}_{k\sigma}}{\tau_{loc}},
\label{eqn:system}
\end{eqnarray}
where $T_{SOI}^{X}$ and $T_{el-phon}^{X}$ are the scattering times due to the spin-orbital interaction and electron-phonon coupling, correspondingly. Parameters $\alpha=\textrm{exp}(-\frac{\Delta E_{B-sf}}{k_{B}T})$ and $\beta=\textrm{exp}(-\frac{\Delta E_{B-mf}}{k_{B}T})$ with $\Delta E_{i}$ being the energy difference between bright exciton and corresponding dark exciton, $k_B$ being the Boltzmann constant, and $T$ the temperature. While typical times for scattering due to the spin-orbital interaction are about the times specific for scattering due to the exchange interaction \cite{Jiang2021}, the electron-phonon scattering times were found to be much faster (about 50-80 fs) \cite{Bertoni2016,Liu2020}. $\tau_{sf}$ and $\tau_{mf}$ are the dark excitons lifetimes, which are typically much longer than the lifetimes of bright excitons. A comparison of PL circular polarization time evolution for bright excitons and trions under linearly polarized light excitation in the case when only intervalley scattering of bright excitons is considered (solid lines) and when both intervalley and intravalley scattering occurs (dashed lines) involving dark excitons is shown in Fig.\ref{fig:comparison}. One can clearly see only quantitative difference between the obtained results.

\section{Conclusions}

We propose a general theoretical model which allows theoretical analysis of valley pseudospin control in the TMD monolayer by utilizing both the
magnetic proximity effect of magnetic monolayer and intervalley scattering of excitons and trions. Within the proposed model we performed a careful analysis of a
substantial difference in exciton and trion polarization under different polarized laser excitations due to the interplay between interlayer charge transfer and intervalley quasi-particles scattering. We demonstrated that the formation of spin polarization in semiconductor monolayer and consequently circularly polarized PL is due to the competition between interlayer charge transfer and intervalley quasi-particles scattering. A mechanism for switching PL polarization sign in TMD/magnetic heterostructure under circularly polarized laser excitations is discussed. We also discussed generalization of the proposed model due to the careful analysis of both intervalley and intravalley scattering processes between bright and dark excitons. An important moment is that the proposed model can be effectively applied to a wide class of van der Waals heterostructures. This means that a long-distance manipulation of exciton and trion behaviors in multilayer heterostructures can be achieved through spin-selective charge transfer. Obtained result marks a significant advancement in the control of spin and valley pseudospin in multilayer structures.

\section{Acknowledgements}
This  work  has  been carried out under the financial support from the
Russian Science Foundation project 25-12-00093 (Model formulation, analysis of exciton and trion PL polarization). V.N.M. also thanks the Foundation
for the Advancement of Theoretical Physics and Mathematic BASIS.




\end{document}